\documentclass[preprint,authoryear,10pt]{elsarticle}

  \usepackage{graphics}
  \usepackage{graphicx}
\usepackage{amssymb}
\begin{document}
  \begin{frontmatter}
 

 
\title{Test of two hypotheses explaining  the size of  populations in a system of cities}


\author{Nikolay K. Vitanov$^{1}$,  Marcel Ausloos$^{2,3,4}$}
 
 \address{
$^1$ Institute of Mechanics, Bulgarian Academy of Sciences,\\
Acad. G. Bonchev Str., Bl. 4, BG-1113 Sofia, Bulgaria \\
$^2$  Royal Netherlands Academy of Arts and Sciences\\
Joan Muyskenweg 25, 1096 CJ Amsterdam, The Netherlands  \\Ê 
$^3$ School of Management, University of Leicester, \\ University Road, Leicester  LE1 7RH, UK  \\  
 $^4$ GRAPES, Beauvallon Res., rue de la Belle Jardiniere, 483/0021\\
B-4031, Liege Angleur, Euroland
 }  
 
\begin{abstract}
Two classical hypotheses are examined about the population  growth in a system of cities {\bf : Hypothesis 1 pertains to Gibrat's  and Zipf's  theory which states that the city growth-decay process   is size independent; Hypothesis 2 pertains to the so called Yule process which states that the growth of populations in cities happens when (i) the distribution of the city  population  initial  size  obeys  a log-normal function, (ii) the growth of the settlements follows a stochastic process.  }
The basis for the test is  some official data on Bulgarian cities at various times. This   system was chosen because  (i) Bulgaria is  a   country  for which one does not expect biased  theoretical conditions;  (ii) the  city  populations were determined rather precisely.
The present results show that: (i) the population size growth of the  Bulgarian cities is size dependent,  whence    Hypothesis  1   is   not   confirmed for  Bulgaria;  (ii) the population size growth of Bulgarian cities can be  described by a double Pareto log-normal distribution, whence Hypothesis  2  is valid for the Bulgarian city system.  It is expected that this fine study brings some information and light on other usually considered to be more pertinent countries of city systems.
\end{abstract}
  \end{frontmatter}

%\begin{keyword}
%City size \sep Gibrat's law  \sep Yule stochastic seed process \sep double Pareto lognormal distribution \sep  Zipf's law \sep Bulgarian cities\\
%{\bf JEL code:  R120} - Size and spatial distributions of regional economic activity
%\end{keyword}
% \linenumbers

%\section{}
%\label{}

\section{Introduction}

The rapid development of the methods of  nonlinear dynamics and those for studying
  time series has led to many applications of  
new and classic methodology to the problems of science, society and economics.
%(see also Appendix A). 
A large number of  applications is devoted to
  nonlinear problems of  economic geography (\cite{a2,a1}). Below we shall discuss
 characteristics that seem important for  understanding the evolution of  complex economic
or social systems of a region, country or a system of countries, i.e.    the
population  size  of   cities in a well defined  geographic area.

Let us consider a city. In order to explain the size of the population
of the city, we have to account for its geographic location, economic 
development, etc. The evolution of a city population   is affected by many factors, 
e.g.,  impulses generated  by the city and its  hinterland,  interurban 
dependencies,  or "shocks" from outside the city system (\cite{k1}). 
Let us turn  now to a system of cities.  As in the case of a single city, the economic,
geographic and many additional factors are also important for the growth  (or decay) of the
populations  of city systems. As  a city system is  distributed over
a geographic region, there are variations in the above factors. These variations
can be modeled by stochastic processes. Thus,  if we are interested  in a
city population size distribution (\cite{Cordoba1,Cordoba2}) the building of a theory can start from appropriate
stochastic processes rather than from model equations for the economic, geographic and
other factors. In other words, in order to explain the distribution
of the city population sizes in a geographic region of a city system it seems that
  a mathematical model based on stochastic processes with
appropriate characteristics is in order (\cite{seto,vx10,cit1,vesp}).

In the course of  time, the cities in a country  develop a  hierarchy. An 
expression of this hierarchy is the city population size $distribution$ that can 
be easily   constructed for any urban system. Zipf (\cite{z1,Ioan})  suggested that a 
large number of observed city population  size distributions could be 
approximated by a simple  {\it scaling (power) law} 
\begin{equation}\label{Zipfeq}
N_r = \frac{C}{r^\beta},
\end{equation}
where $N_r$ is the population of the $r$-th largest city, {\bf where $C$ is  some "constant", which value is obviously constrained by  a normalization condition},  and $\beta=1$.
Eq.(\ref{Zipfeq}) is called the rank-size scaling law.  Zipf suggested that 
the particular case  $\beta=1$ represents a desirable situation (rank-size rule),
in which the  forces of concentration balance those of decentralization. 
It has been observed that  the  urban population size distributions  in developed countries, like 
the USA,  fits very well the rank-size rule, over several decades (\cite{k1,madden}).  {\bf For broadening the view on possible other cases of interest, i.e. where a power law, as in Eq. (\ref{Zipfeq}),  is found, let us mention  work by    \cite{JangTia11} on USA,   \cite{kt2} on Denmark, \cite{mora06} on Brazil,  \cite{gangbas} on India and China,  and  \cite{Peng010} on China.}
  % Note that the distribution of population in cities can be  also described by using a $q$-exponential distribution (\cite{0109232_cities}).

In this paper,  two hypotheses are tested about the growth of a population in a system
of cities. For the test,  we have selected the city system of Bulgaria. One 
reason   stems from the fact  that the population of Bulgarian cities can be determined very 
precisely; in particular,  the "city" is well defined: there is no suburban area,  as  often found in the large  or populated countries such as,  for example, the USA, India, China, France, or  Italy. 

The data sets consists of the yearly count of the  population of  whole Bulgarian cities  from 2004 till 2011,  as  recorded by 
the National Statistical Institute of the Republic of Bulgaria  
($http: www.nsi.bg$). 

Thus, the hypotheses to be discussed are : 
\begin{itemize}
\item
\textbf{Hypothesis 1:} \textit{The growth rate of the "rescaled city population" 
is independent on the  size of the population.}
\item
\textbf{Hypothesis 2:} \textit{The settlement formation follows a  Yule process, in which
 the initial populations of the settlements  are distributed according to the log-normal
function. The evolution of the city populations of the formed settlements 
follows a stochastic process, like the geometric Brownian motion.}
\end{itemize}
 \textbf{Hypothesis 1} is a formulation of the Gibrat's law which leads to 
power law distributions of the rescaled sizes of a system of cities. % see also  Appendix B.  
 \textbf{Hypothesis 2} gives the necessary conditions for 
describing the distribution of the (non-rescaled) city population sizes of a system of
cities {\bf along}  the  double Pareto log-normal distribution. %  see also Appendix C. 
\section{Test of  \textbf{Hypothesis 1}} \label{sect2}

The $i$-th city   rescaled size is  defined as
 \begin{equation} \label{Seq}
S_i(t) = N_i(t)/N(t), 
\end{equation}
 at time $t$, where $N_i(t)$ is the population of the $i$-th city and $N(t)$ is the population of all cities in year $t$.    The (rescaled city) sizes are  ranked such that $r=1, 2, ...$ with $N_r(t)/N(t) \ge N_{r+1}(t)/N(t)$. In so doing, the (rescaled) rank ($r$)-size relationship reads
\begin{equation}\label{lnSeq}
 \ln(r)= \alpha - \beta \ln(S)  
 \end{equation}
where $\alpha$ and $\beta$ are parameters.
\begin{figure}[t]
 \vskip1cm
\begin{center}
\includegraphics[scale=0.8]{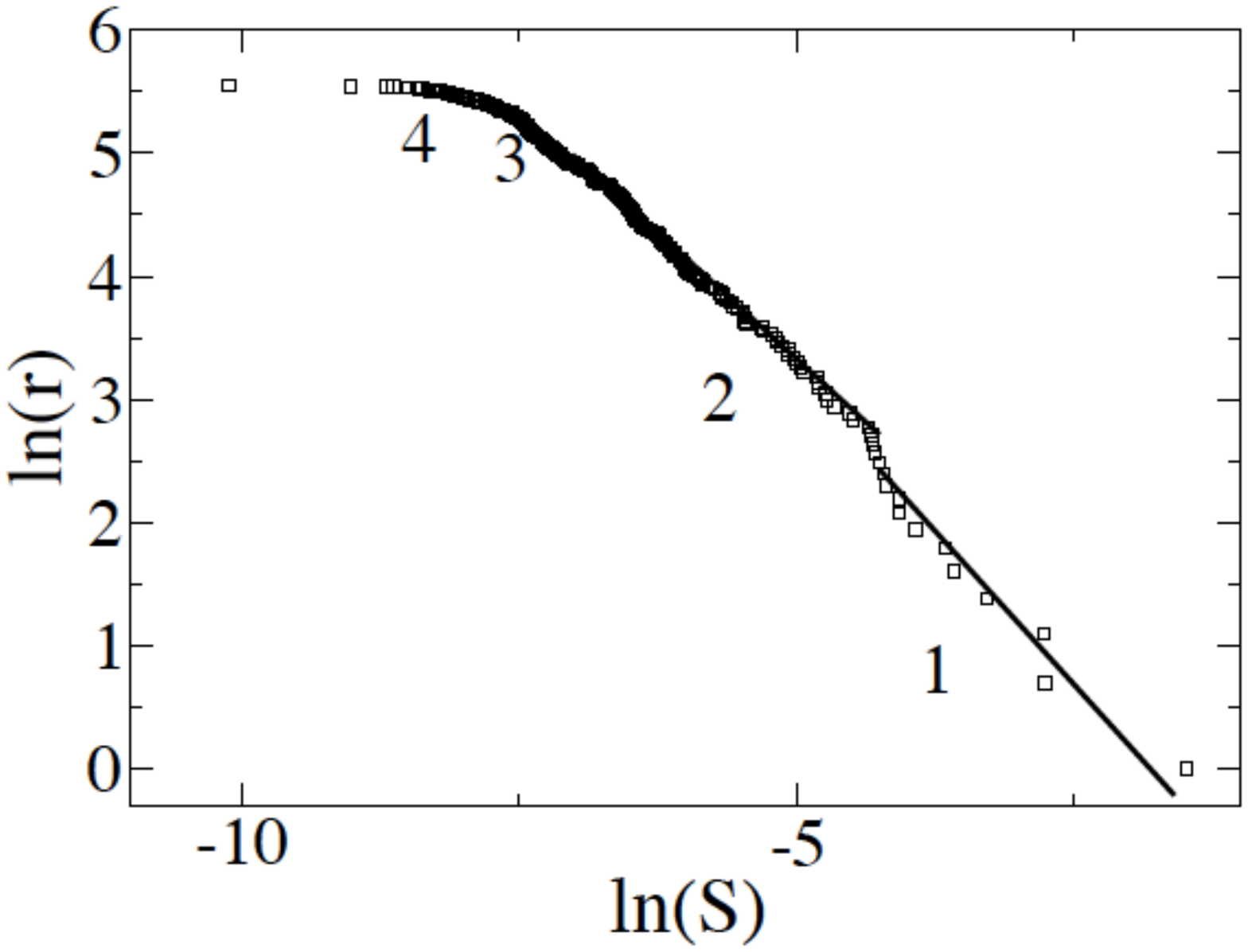}
\caption{ .}
\label{Fig1} \end{center} \end{figure}

After {\bf visual inspection and} much statistical analysis of the data, {\bf i.e., adapting   the limits of the relevant intervals by trial and error procedures}, it is found that the cities  are best grouped into four classes: large cities (class 1); medium size cities (class 2); small cities (class 3) and very small cities  (class 4).   {\bf A comment on the various sizes has to be found below, after  the data  specific analysis.}  Fig. \ref{Fig1} outlines   the rank-size relationship  for the rescaled  sizes $S$ of the system of Bulgarian cities,    in 2004. {\bf The  figures for other analyzed years, 2005-2011,  are not shown, for conciseness, but are very similar to the 2004 case.}
\begin{table}[t]
\begin{center}
\begin{tabular}{|l|l|l|l|l||l| }
  \hline 
  \multicolumn{6}{|c|}{2004 } \\ \hline
\hline
[Pop]&$N_c$  & Cl. & $\alpha$ & $\beta$ & $\gamma$ \\
\hline
\hline
1.14 10$^6$&16 &  1 & $-1.79 \pm 0.19$ & $ 0.99 \pm 0.05$ & - \\
\hline
70 000&122 &   2 & $ -0.77 \pm 0.03$& $ 0.820 \pm 0.004$ & - \\
\hline
5 000&93&   3& $-20.99 \pm 0.14$ & $6.47 \pm 0.21$ & $0.39 \pm 0.02$ \\
\hline
2 000&17 &   4 & $4.42 \pm 0.06$ & $0.128 \pm 0.007$ & - \\
  \hline  \hline 
  \multicolumn{6}{|c|}{2011 } \\ \hline
[Pop]&$N_c$ & Cl. & $\alpha$ & $\beta$ & $\gamma$ \\
\hline
\hline
1.21 10$^6$&15 &   1& $-1.46 \pm 0.18$ & $0.89 \pm 0.06$ & - \\
\hline
70 000&120&    2& $-0.910 \pm 0.002$ & $0.837 \pm 0.004$& - \\
\hline
5 000&94 &   3& $-23.94 \pm 0.19$& $ 7.16 \pm 0.24$& $ 0.43 \pm 0.03$ \\
\hline
2 000&28&   4& $ 3.82 \pm 0.07$ & $ 0.204 \pm 0.008$& - \\
%\noalign{\smallskip}
\hline
\end{tabular} \caption{Values of the parameters $\alpha$, $\beta$, and $\gamma$ for the
four classes of Bulgarian cities for 2004 and 2011; see Eqs.(\ref{lnSeq})-(\ref{lnS2eq}). {\bf The number of cities in each Class (Cl.) is given together   with the respective population upper size ([Pop]) in the Class.}}\label{Table1}
\end{center}
\end{table}

It is found  that
the rank-size relationship for the classes 1, 2, and 4
can be approximated by a straight line (Zipf's law), Eq.(\ref{lnSeq}).
The parameters $\alpha$ and $\beta$ for these classes are given in Table 1.
Note that for the case of large cities the exponent  
$\beta$ is almost $1$, as conjectured by Zipf to be an optimal  case. 
In contrast, the rank-size relationship for the  small cities  in class 3
is very well  approximated  by a $quadratic$ regression of the kind 
\begin{equation}\label{lnS2eq}
 \ln(r)= \alpha - \beta \ln(S) - \gamma (\ln(S))^2, 
 \end{equation}
where the corresponding parameters are written in Table \ref{Table1}.
For comparison, the parameters for the 
rank-size distribution of the Bulgarian 
cities for 2011 are also given in Table \ref{Table1}.

 {\bf For completeness, the number of cities and their population ranges in each class are given in Table \ref{Table1}. We emphasize that the grouping is not made {\it a priori}  but results from data analysis. The fact that the analysis of unbiased  measurement  yields consistent results over time  allows us to state that the prediction can be said to be
reliable.

We  only show 2004 and 2011, but indicate in the text, that these are only examples which are confirmed in other years. On the basis of above we have reliability of the hypothesis for the entire studied interval from 2004 till 2011. } This grouping is not unusual as such a separation into classes was observed also for the case of other  countries, cf., e.g., (\cite{27,47}).

Thus, Fig. \ref{Fig1} and Table \ref{Table1} indicate that   \textbf{Hypothesis 1} is not valid for the Bulgarian city system. 

Let us supply  some more
argument for such a conclusion, e.g. through a Pareto plot (\cite{Pareto,r1,r2}),  i.e. let it be searched whether the cumulative distribution function  of such a population system  follows an inverse power law of $S$.

\begin{figure}[t]
\begin{center}
\includegraphics[scale=0.8]{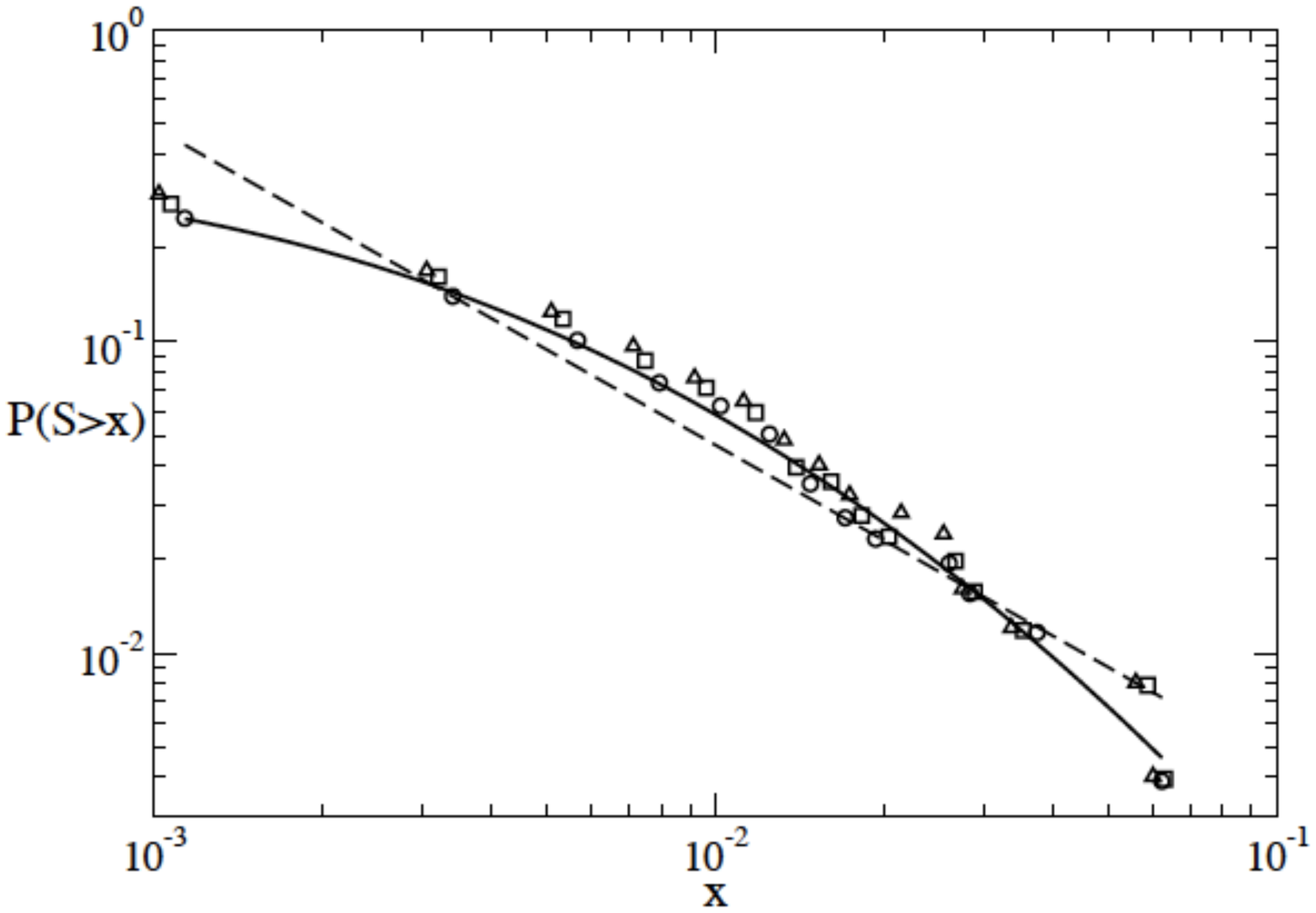}
\end{center}
\caption{ }  \label{Fig2}
\end{figure}

 The theory of the size-independent growth
%(see Eq.(\ref{aa6}) from Appendix A) 
predicts a power law for $P(S>x) \sim S^{-\zeta}$, with 
a constant  $\zeta$.  The probability $P(S>x)$ that the rescaled size of a  Bulgarian city  population
exceeds a certain value $x$ in 2011 is shown in Fig. \ref{Fig2}. In Fig. \ref{Fig2}, this theory is represented by
the dashed line,  i.e.  a power law fit to the data. It is easily seen that a power law
%as that of Eq.(\ref{aa6})
 is not a good approximation for the (rescaled) city  population size distribution in  Bulgaria. An extended version of the theory of the city growth
should allow for a  population size dependent growth. As a consequence of this, the exponent $\zeta$ depends on the population size of the city.
% (see Eq.(\ref{aa9}) from Appendix A).  
 Therefore,   some $\zeta(x)$ can be estimated as follows.

\begin{figure}[t]
\begin{center}
\includegraphics[scale=0.8]{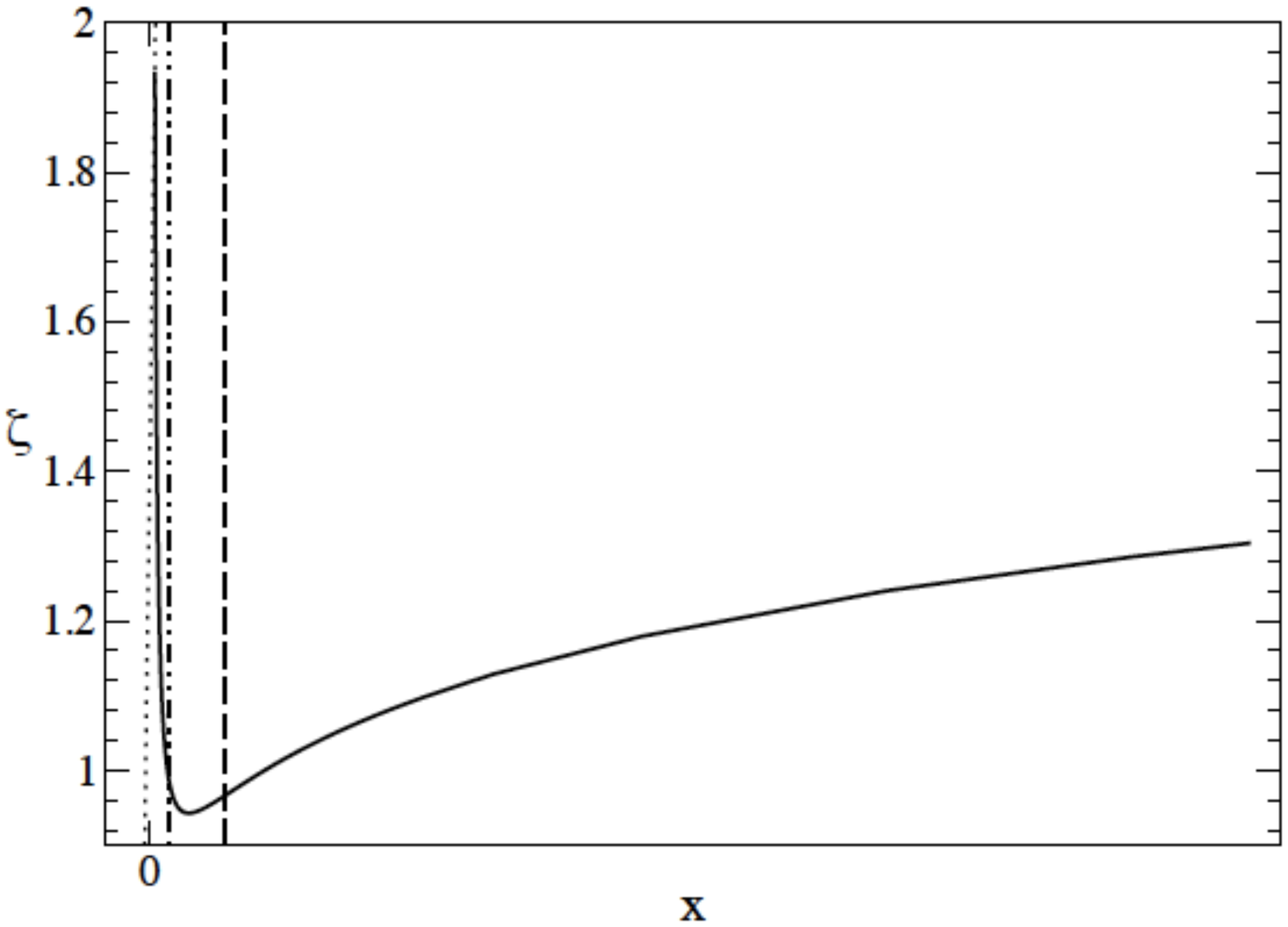}
\end{center}
\caption{ } \label{Fig3}
\end{figure}

The solid line in Fig. \ref{Fig2} represents the fit of the probability $P(S>x)$
for the system of Bulgarian cities for 2011. This fit corresponds to the
distribution
\begin{equation}\label{d1}
P(x) = \frac{a}{2} {\rm erfc} \left[ - \frac{\ln (x/b)}{c \sqrt{2}} \right]
\end{equation} 
where $a$, $b$ and $c$ are parameters and ${\rm erfc}(x) = 1 - {\rm erf(x)} $
where ${\rm erf (x)}$ is   %given by  Eq. (\ref{ab7}) in Appendix B. 
\begin{equation}\label{ab7}
{\rm erf}(x) = \frac{2}{\sqrt{\pi}} \int_{0}^x dy \ e^{-y^2}.
\end{equation} 
In order to
obtain $\zeta(x)$,  the theoretical distribution
$(x/S_{min})^{-\zeta(x)}$ is  assumed to be the same as Eq.(\ref{d1}), i.e.,
\begin{equation}\label{d2}
\left(\frac{x}{S_{min}} \right)^{-\zeta(x)} = \frac{a}{2} {\rm erfc} 
\left[ - \frac{\ln (x/b)}{c \sqrt{2}} \right].
\end{equation}
From Eq.(\ref{d2}),   the $local$ exponent $\zeta(x)$ is so obtained: 
\begin{equation}\label{d3}
\zeta(x) = - \frac{\ln(a/2) + \ln \{ {\rm erfc} [-(\ln(x/b))/(c \sqrt{2})] \}}{\ln (x/S_{min})}.
\end{equation}
The local exponent from Eq.(\ref{d3}) is shown in Fig. \ref{Fig3}.

For a large interval of $x$ the value of $\zeta$ is close to $1$, the Zipf value.
Large deviations from $1$ are observed for the class of small cities. 
If  \textbf{Hypothesis 1} was true for the  whole system of 
Bulgarian cities,  $\zeta$ should be a constant corresponding to a   straight line  in
  Fig.  \ref{Fig3}. This is not the case,{\bf thereby } implying  that 
\textbf{Hypothesis 1} is not valid for the system of Bulgarian cities. 
%Thus in general the growth of the city population is dependent on the size of the  population.
%
\section{Test of \textbf{Hypothesis 2}}
It is emphasized first that the \textbf{Hypothesis 2} is formulated for the non-rescaled
sizes of the populations in a system of cities. The test of the hypothesis
goes as follows:  the city population size distribution for the Bulgarian
cities  is fitted to the double Pareto log-normal distribution,  %Eq.(\ref{ab6}).
\begin{eqnarray}\label{ab6}
p(x) = \frac{\alpha \beta}{2(\alpha + \beta)} \exp \left(\alpha \mu +
\frac{\alpha^2 \sigma^2}{2} \right) x^{-\alpha -1} \left[ 1 + {\rm erf} \left(  
\frac{\ln(x) - \mu - \alpha \sigma^2}{\sigma \sqrt{2}} \right) \right] 
+ \nonumber \\
\frac{\alpha \beta}{2(\alpha + \beta)} \exp \left( -\beta \mu +
\frac{\beta^2 \sigma^2}{2} \right) x^{\beta -1} \left[ 1 - {\rm erf} \left(  
\frac{\ln(x) - \mu + \beta \sigma^2}{\sigma \sqrt{2}} \right) \right]
\nonumber \\
\end{eqnarray}
If the assumptions of \textbf{Hypothesis 2} are correct,  it  should be expected 
that the double Pareto log-normal distribution  provides a good fit to the
data.  However, if one or more of the assumptions in \textbf{Hypothesis 2}
are not valid,   the fit  should not be a satisfactory one.

\begin{figure}[t]
\begin{center}
\includegraphics[scale=0.8]{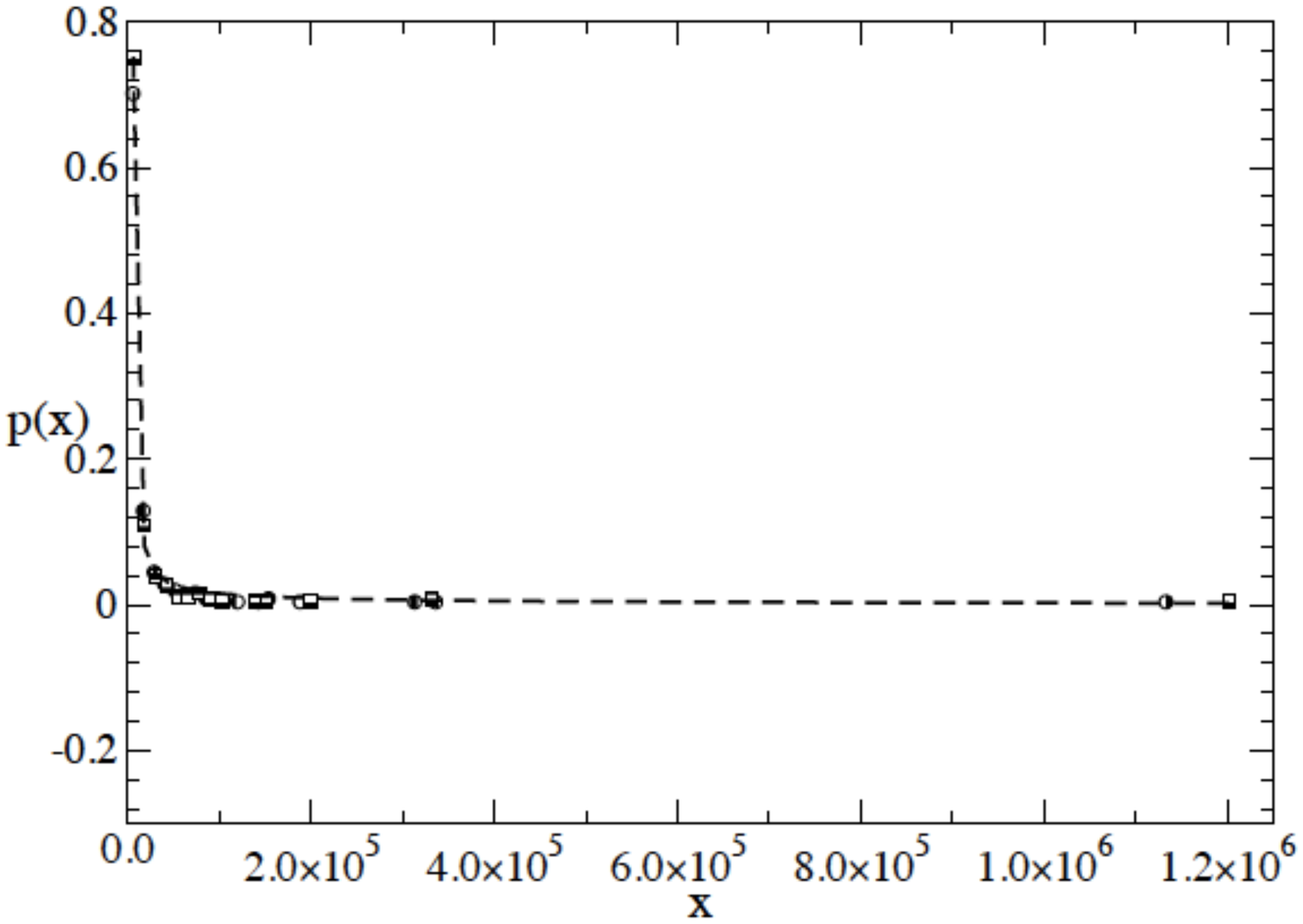}
\end{center}
\caption{}  \label{Fig4}.
\end{figure}

The distribution of population sizes of the 
system of Bulgarian cities in 2004 and in 2011 are presented  in Fig. \ref{Fig4}.  
The dash line represents  the double Pareto log-normal distribution that
corresponds to the distribution of the Bulgarian cities in 2011.  From the
values of $\alpha$ and $\beta$ and from the properties of the double Pareto
log-normal distribution it follows that $p(x)
\approx x^{-0.8}$ , for the case of small city sizes and $p(x) \approx x^{-3}$ for the large cities. Similar results are found for the years for which the data are available.

Thus,  the double Pareto log-normal distribution fits well
the city population size distributions for the Bulgarian city system  whatever the year in this century. Thus  \textbf{Hypothesis 2} is
confirmed.
\section{Concluding remarks}
    
The study of the size of populations in cities of a
country has many economic applications. Apart on the decision taking about
placing industrial plants or building trade centers, the city population sizes influence
the efficient use of   resources or the possibilities of economic growth 
(\cite{conc1,conc2}). Thus, the investigation of the evolution of the
populations in a system of cities is an important research topic. 

 Our study  reported here 
above has shown that for the case of the Bulgarian city system 
\textbf{Hypothesis 1} is not confirmed but  \textbf{Hypothesis 2} is confirmed.
The validity of   \textbf{Hypothesis 2} even gives some information about the
history of  settlement formations in Bulgaria. It seems that the distribution
of the initial populations of the settlements was indeed log-normal and that
the formation happened according to the Yule process, i.e. the new settlements have
been formed on the basis of the existing older settlements nearby. 
\begin{flushleft}
\textbf{Acknowledgment}
\end{flushleft}
  {\small This work has been performed in the framework of COST Action IS1104 
"The EU in the new economic complex geography: models, tools and policy evaluation".
We acknowledge some support
through the project 'Evolution spatiale et temporelle d'infrastructures r\'egionales 
et \' economiques en Bulgarie et en F{\'e}d{\'e}ration Wallonie-Bruxelles' 
within the 
intergovernmental agreement for cooperation between the Republic of  Bulgaria and la Communaut\' e 
Fran\c{c}aise de Belgique.}
\newpage

{\bf Supplemental  material }
\vskip0.5cm
Supplemental material can be found containing  some brief review of the pertinent literature on nonlinear dynamics, nonlinear time series analysis and their
applications. The two hypotheses tested in the
main text of the paper being closely connected to the existence of power laws
for the population sizes of the cities  are discussed along analytical lines. It is  shown that Kesten  	and Gibrat  proportional random growth processes
lead to power law distributions. On the other hand, the Yule process of settlement formation and the  geometric Brownian motion  assumed for their growth is shown to result in a  double Pareto log-normal distribution.
\vskip05cm

\newpage 
\vskip0.5cm

Fig. 1  Rank-size relationship for the system of Bulgarian cities   in  2004. 
The cities are divided into 4 classes. The rank-size relationships for the 
classes 1, 2, and 4 are fitted  by a linear regression $\ln(r) = \alpha - \beta \ln(S)$. 

\vskip0.5cm

Fig. 2      Probability $P(S>x)$
that the rescaled population city size $S_i$ of Bulgarian cities is larger than some value $x$ in  2004 (triangles),    2007 (squares), and  2011 (circles). Dashed line:
Power law distribution with a constant exponent  $\zeta =
1.02$. % S_{min}=4.96\cdot 10^{-4}$. 
Solid line:  Eq.(\ref{d1}) with $a=0.359$; $b=0.0023$;
$c=-1.469$.

\vskip0.5cm

Fig. 3  Local exponent $\zeta(x)$ from Eq.(\ref{d3}) with parameter values  of 
%in Eq.(\ref{d3}) are the same as in 
Fig.2. %namely: $S_{min}=4.96\cdot 10^{-4}$;  $a=0.359$; $b=0.0023$; $c=-1.469$. 
The vertical lines mark the boundaries
between the  4 classes of cities from large to small sizes with increasing $x$. % as follows. To the right from the vertical dashed line : values of $\zeta(x)$ for the class of large cities. Between the vertical double dot-dashed line and the vertical dashed line: values of $\zeta(x)$ for the class of the medium sized cities. Between the vertical doted line and the vertical double dot - dashed line: values of $\zeta(x)$ for the class of small cities. To the left from the vertical dotted line: values of $\zeta(x)$ for the class of very small cities.  

 \vskip0.5cm

Fig. 4  Probability density functions for  Bulgarian city population size in 2004
(circles) and 2011 (squares)  with  a  double Pareto log-normal distribution 
fit   Eq.(\ref{ab6}) 
(dashed line) % of the p.d.f.  of the cities 
for 2011. The
parameters of the double Pareto log-normal distribution from the dashed line are:
$\alpha = 2.049$; $\beta = 0.216$, $\sigma = -6.188$; $\mu = -25.737$. %For the city size p.d.f. 
 $R^2=0.996$ in  2011. 
\end{document}